\documentclass[letter,longauth]{aa} 
\usepackage{graphicx}
\usepackage{natbib}
\bibpunct{(}{)}{,}{a}{}{;}
\usepackage{amssymb}
\usepackage{amsmath}
\usepackage{txfonts}

\newcommand{\trot}{\mbox{$T_{Rot}$}}

\begin{document}
\titlerunning{DIGIT: PACS Spectral Observations of DK Cha}
\title{Dust, Ice and Gas in Time (DIGIT) Herschel\thanks{Herschel is an ESA space observatory with science instruments
provided by the European-led Principal Investigator consortia and
with important participation from NASA.} program first results: 
}
\subtitle{A full PACS-SED scan of the gas line emission in protostar DK Cha.}
\author{ T.~A.~van~Kempen\inst{1}
\and J.~D.~ Green \inst{2}
\and N.J.~Evans~II\inst{2}
\and E.F.~van~Dishoeck\inst{3,4}
\and L.E.~Kristensen \inst{3}
\and G.~J.~Herczeg \inst{4}
\and B.~Mer\'in\inst{5}
\and J.-E.~Lee\inst{6}
\and J.~K.~J{\o}rgensen\inst{7}
\and J.~Bouwman\inst{8}
\and B.~Acke \inst{9,24} 
\and M.~Adamkovics \inst{10} 
\and J.C.~Augereau\inst{11} 
\and E.~Bergin\inst{12} 
\and G.~A.~Blake\inst{13} 
\and J.~M.~Brown \inst{4} 
\and J.~S.~Carr\inst{14} 
\and J.-H.~Chen\inst{2}
\and L.~Cieza\inst{15} 
\and C.~Dominik\inst{16,17}
\and C.~P.~Dullemond\inst{8,18}
\and M.~M.~Dunham\inst{2}
\and A.~Glassgold\inst{10}
\and M.~G\"udel\inst{19}
\and P.~M.~Harvey\inst{2}
\and Th.~Henning\inst{8}
\and M.~R.~Hogerheijde\inst{3}
\and D.~Jaffe\inst{2}
\and H.~J.~Kim\inst{2}
\and C.~Knez\inst{20}
\and J.~H.~Lacy\inst{2}
\and S.~Maret\inst{11}
\and G.~Meeus \inst{21}
\and R.~Meijerink \inst{3}
\and G.D.~Mulders\inst{16,22}
\and L.~Mundy\inst{20}
\and J.~Najita\inst{23}
\and J.~Olofsson\inst{8}
\and K.~M.~Pontoppidan\inst{13}
\and C.~Salyk\inst{2}
\and B.~Sturm\inst{8}
\and R.~Visser\inst{3}
\and L.B.F.M.~Waters\inst{9,16}
\and C.~Waelkens\inst{9}
\and U.A.~Y{\i}ld{\i}z\inst{3}
      }

\institute{
Harvard-Smithsonian Center for Astrophysics, 60 Garden Street, MS 78,
Cambridge, MA 02138, USA. e-mail: \\
\and
The University of Texas at Austin, Department of Astronomy,
1 University Station C1400, Austin, Texas 78712-0259, USA
\and
Leiden Observatory, Leiden University, PO Box 9513, 2300 RA Leiden,
The Netherlands
\and
Max-Planck-Institut f\"ur extraterrestriche Physik. Garching, Germany
\and
Herschel Science Centre, European Space Astronomy Centre (ESA),
P.O. Box 78, 28691 Villanueva de la Ca\~nada (Madrid), Spain
\and
Department of Astronomy and Space Science, Astrophysical Research
Center for the Structure and Evolution of the Cosmos, Sejong University,
Seoul 143-747, Republic of Korea
\and
Centre for Star and Planet Formation, Natural History Museum of Denmark,
University of Copenhagen, {\O}ster Voldgade  5-7, DK-1350 Copenhagen K.
\and
Max Planck Institute for Astronomy, K\"onigstuhl 17, D-69117
Heidelberg, Germany
\and
Instituut voor Sterrenkunde, K.U.Leuven, Celestijnenlaan 200D, B-3001
Leuven, Belgium
\and
Astronomy Department, University of California, Berkeley, CA 94720, USA
\and
Laboratoire d'Astrophysique de Grenoble, CNRS/Universit\'{e}
Joseph Fourier (UMR5571) BP 53, F-38041 Grenoble cedex 9, France
\and
Department of Astronomy, The University of Michigan, 500 Church Street,
Ann Arbor, MI 48109-1042, USA
\and
Caltech, Division of Geological \& Planetary Sciences, Mail Code 150-21,
Pasadena, CA 91125
\and
Naval Research Laboratory, Code 7211, Washington, DC 20375, USA
\and
Institute for Astronomy, University of Hawaii at Manoa, Honolulu,
HI 96822, USA ; Spitzer Fellow
\and
Astronomical Institute ``Anton Pannekoek'', University of Amsterdam,
PO Box 94249, 1090 GE Amsterdam, The Netherlands
\and
Department of Astrophysics/IMAPP, Radboud University Nijmegen,
P.O. Box 9010 6500 GL Nijmegen The Netherlands
\and
Institut f\"ur Theoretische Astrophysik, Universit°t Heidelberg,
Albert-Ueberle-Strasse 2, 69120 Heidelberg, Germany
\and
University of Vienna, Department of Astronomy, T\"urkenschanzstr. 17,
1180 Vienna, Austria
\and
Department of Astronomy, University of Maryland, College Park,
MD 20742, USA
\and
Dpt. de F\'{i}sica Te\'{o}rica, Fac. de Ciencias, Universidad
Aut\'{o}noma de Madrid, Cantoblanco, 28049 Madrid, Spain
\and
SRON Netherlands Institute for Space Research, PO Box 800, 9700 AV,
Groningen, The Netherlands
\and
National Optical Astronomy Observatory, 950 N. Cherry Ave.,
Tucson, AZ 85719, USA
\and
\fnmsep\thanks{Postdoctoral Fellow of the Fund for Scientific Research,
Flanders}
}

\offprints{T.A. van Kempen : tvankempen@cfa.harvard.edu}

\date{Draft: 20 April 2010}


\def\placeFigureSED{
\begin{figure}[!th]
\begin{center}
\includegraphics[width=240pt]{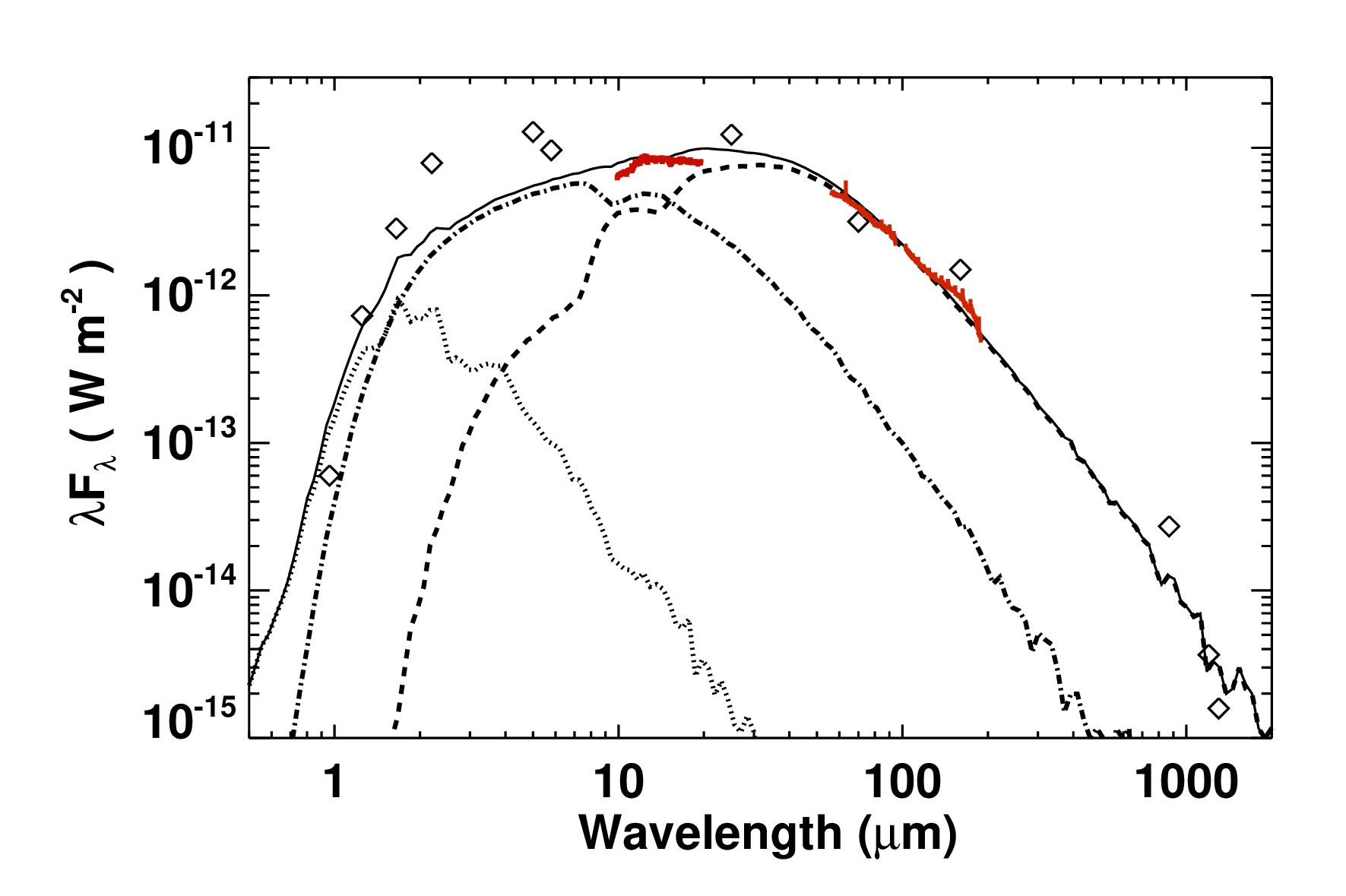}
\end{center}
\caption{SED of DK Cha, using WFI I-band data, 2MASS (J,H,K), 
Spitzer-IRAC  (3.6 and 4.5 $\mu$m. The other bands, 5.8 and 8.0 $\mu$m are not shown due to saturation), 
Spitzer-MIPS (24, 70 and 160 $\mu$m), APEX-Laboca (870 $\mu$m) and 
SEST (1.2 and 1.3 mm).  Overplotted are the best-fit model from the
\citet{Robitaille07} grid and the PACS and Spitzer-IRS spectra. The MIPS 70 $\mu$m is corrected for saturation; the 160 $\mu$m for large-scale emission.  The solid line is the full model,
while the dashed line shows the envelope contribution, the dot-dashed the
disk contribution, and the dotted the star. Photometric points are shown
as diamonds. The short-wavelength points are not well matched.
Fluxes at short wavelength ($<$3 $\mu$m) are variable.
}

\end{figure}
}

\def\placeFigureLines{
\begin{figure*}[!th]
\begin{center}
\includegraphics[width=450pt]{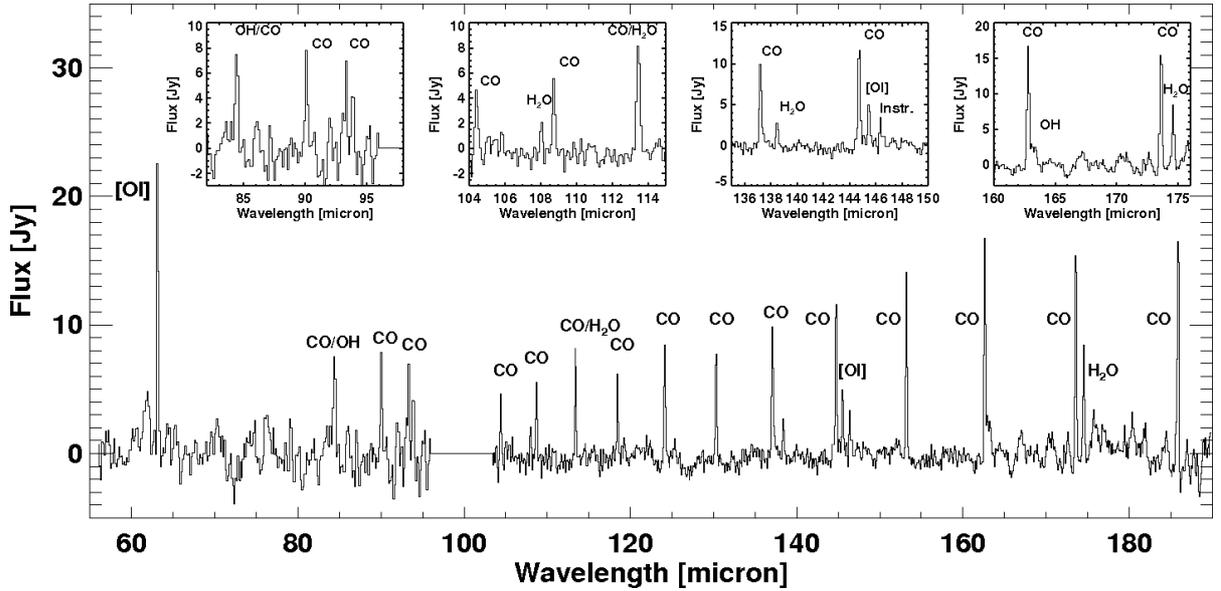}
\end{center}
\caption{The spectrum from 55-190 $\mu$m of all spaxels
added together, after removal of the continuum and
rebinning to 0.5 $\mu$m resolution.
Several dips or bumps (e.g., around 75 $\mu$m) are instrument artifacts.
Inserts show expanded views of parts of the spectrum, with line
identifications.
}
\label{fig:lines}
\end{figure*}
}

\def\placeFigureCO{
\begin{figure}[!th]
\begin{center}
\includegraphics[width=190pt]{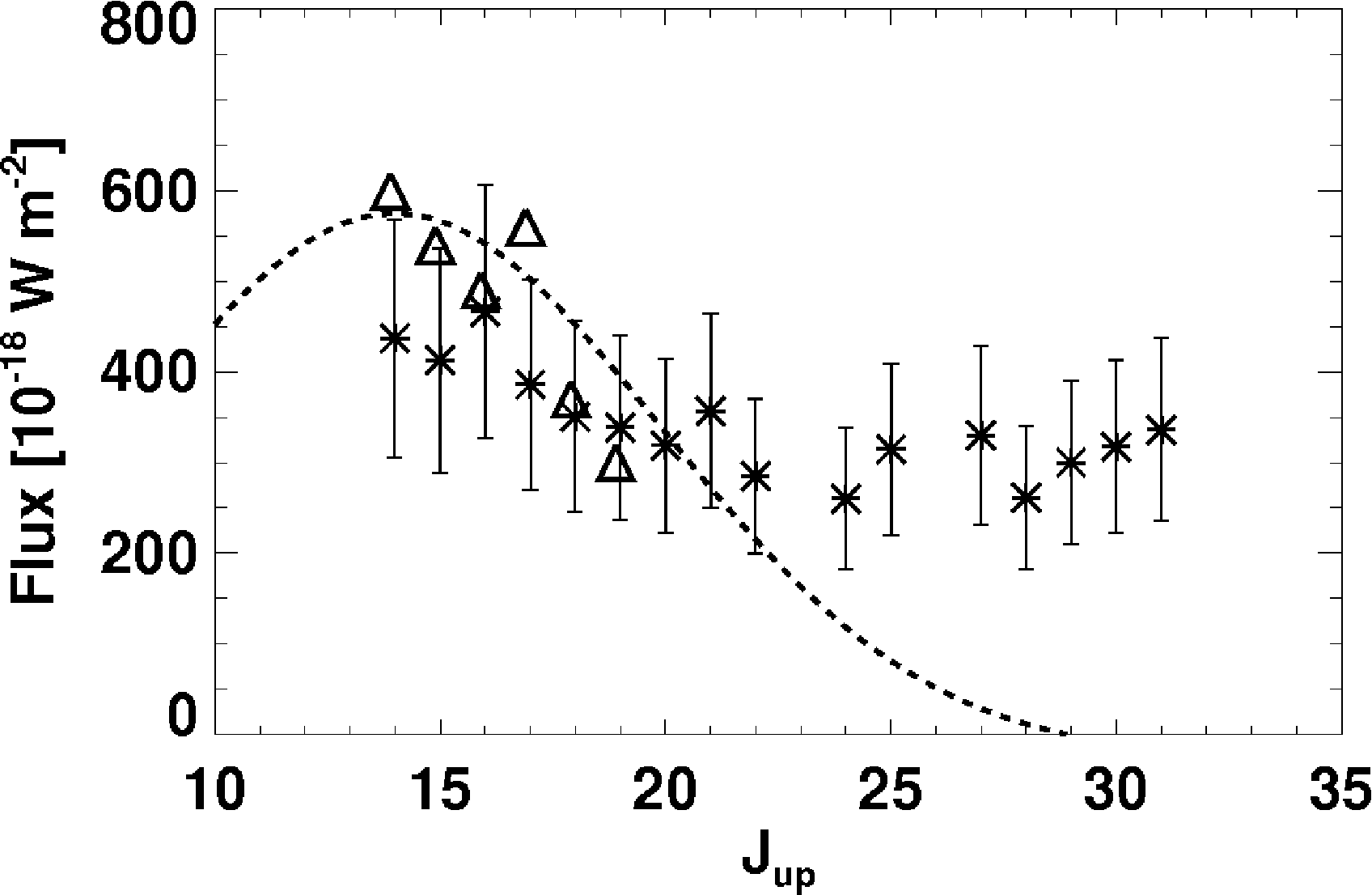}
\includegraphics[width=190pt]{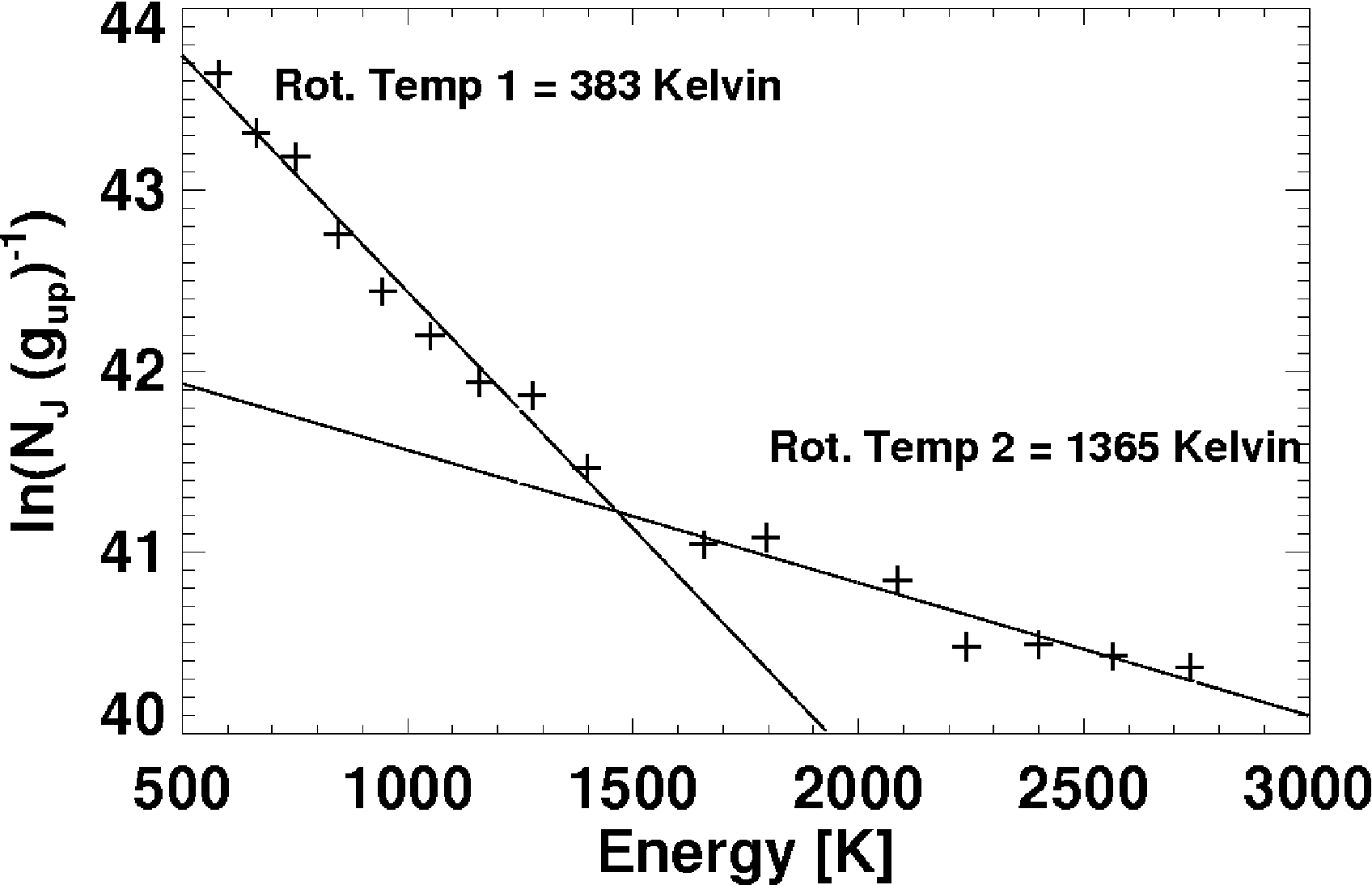}
\end{center}

\caption{{\it Upper:} CO line fluxes plotted versus $J_{\rm{up}}$.
The ISO line fluxes are shown with triangles.
A clear excess can be seen in the CO line fluxes from $J_{\rm{up}}=$20
onwards compared to the warmest model of 750 K proposed by \citet{Giannini99}
(dashed line). 
{\it Lower:} CO Rotational diagram,
showing two distinct populations with a break around 1500 K in energy.  The blended lines are excluded from the fit and the plot.
CO lines with $J_{\rm{up}}$ larger than 31 are excluded from both plots, 
as these lines are detected only in the central spaxel.}
\end{figure}
}



\def\placeTableOne{
\begin{table}[!th]
\caption{Complementary photometry}
\tiny
\begin{center}
\begin{tabular}{l l l l l}
\hline \hline
Wavelength & Origin & Beam/Aperture & Flux & Ref.\\
(micron) & & (arcsec) & Jy & \\ \hline
0.44  & NOMAD & & 1e-5 & 1 \\
0.55 & NOMAD & & 1.4e-4 & 1 \\
0.64 & WFI & & 8.3e-4 & 2\\
0.79 & WFI & & 6.73e-3 & 2\\
0.96 & WFI & & 1.93e-2 & 2\\
1.25 & 2MASS & & 3.04e-1 & 3 \\
1.65 & 2MASS & & 1.57 & 3 \\
2.2 & 2MASS & & 5.8 & 3 \\
3.4 & CTIO& & 12.32 & 4 \\
4.8 & CTIO& & 21.46 & 4 \\
5.8 & IRAC-3 & & 18.7 & 5\\
8.0 & IRAC-4 & & 11.0 & 5 \\
25.0 & IRAS-25 & & $<$102.8 & 6 \\
70 & MIPS-2 & & $>$37.0 & 5,7 \\
160 & MIPS-3 & & $>$80.0 & 7\\
870 & LABOCA & & 7.92 & 8 \\
1200 & SIMBA & & 1.47 & 9 \\
1300 & SIMBA & & 0.69 & 10 \\ \hline
\end{tabular}\\
\label{tab:phot}
\end{center}
Many of the fluxes can be found in \citet{Alcala08}, but we provide here the
original references: 1: \citep{Zacharias04}, 2:\citep{Spezzi07}, 3: 2MASS
catalogue, 4:\citep{Hughes92}, 5:\citep{Alcala08}, 6:IRAS catalogue, 7: Dunham,
priv. comm, 8:van Kempen, Nefs, in prep, 9:\citep{Young05a}, 10:
\citep{Henning93}
\end{table}
}

\def\placeTablePACS{
\begin{table}[!th]
\caption{Detected gas lines within the PACS spectrum of DK Cha. }
\tiny
\begin{center}
\begin{tabular}{l l r r r r}
\hline \hline
Species & Transition & $\lambda$ $^a$ & E$_u$& FWHM & Flux$^b$  \\
& & $\mu$m  & K & $\mu$m & W m$^{-2}$ \\ \hline
[\ion{O}{i}]  & $^3$P$_1$-$^3$P$_2$  &   63.18  & 227.72  & 0.08   & 3657.7\\
OH 3/2-3/2 &   9/2- - 7/2+ &   65.13  & 512.1 & 0.05 &  56.0$^2$ \\
OH 3/2-3/2 &   9/2+ - 7/2-       &   65.28  & 510.9 & 0.07 &   52.3$^2$\\
CO & 38-37 & 69.07 & 4079.98 & 0.06 & 42.1$^2$ \\
CO & 37-36 & 70.91 & 3871.88 & 0.09 & 59.6$^2$ \\
o-H$_2$O$^3$& 7$_{07}$-6$_{16}$&   71.95  & 843.5 & 0.06 &   37.1$^2$\\
CO     & 36-35             &   72.84  & 3668.96 & 0.06 &   40.1$^2$\\
CO    &  35-34            &   74.89  & 3471.44 & 0.1 &   57.0$^2$\\
o-H$_2$O& 4$_{23}$-3$_{12}$&   78.74 & 432.2  & 0.04 &   19.8$^2$ \\
OH 1/2-3/2  &  1/2- - 3/2+        &   79.12  & 181.7 & 0.06   & 28$^2$\\
CO       &  33-32 & 79.36  & 3092.45 & 0.03   & 17.5$^2$\\
CO      &          32-31   &   81.81  & 2911.15 & 0.05   &  126.2\\
o-H$_2$O  & 6$_{16}$-5$_{05}$ & 82.09 & 643.5   & 0.03  &   96.2\\
CO      &          31-30   &   84.41  & 2735.28 & 0.09 &  336.0 \\
OH   3/2-3/2    &   7/2+ - 5/2-    &   84.42$^1$  &  291.2 & - & -\\
OH  3/2-3/2     &   7/2- - 5/2+     & 84.60  & 290.5 & 0.09  &  301.4\\
CO       &          30-29  &   87.19 & 2564.94 & 0.09  &   317.5\\
CO       &         29-28   &   90.16 & 2399.93 & 0.14  &  545.8\\
CO       &         28-27   &   93.35 & 2240.30 & 0.07  &  261.4\\
CO      &        27-26     &   96.77 & 2086.22 & 0.10   &  330.1\\
CO      &        25-24     &   104.44 & 1794.23 & 0.2     &   314.5\\
o-H$_2$O &   2$_{21}$-1$_{10}$&   108.08 & 194.1 & 0.15     &  126.9\\
CO    &          24-23     &   108.76 & 1656.55 & 0.15     &  260.0\\
CO    &          23-22     &   113.46 & 1524.19  & 0.2 &  477.4\\
o-H$_2$O & 4$_{14}$-3$_{03}$  &  113.54$^1$& 323.5 & - & - \\
CO     &         22-21     &   118.58 & 1397.43 & 0.17    & 285.2\\
OH 3/2-3/2  &  5/2- - 3/2+      &  119.23 & 120.7 & 0.18 &   $<$40\\
OH 3/2-3/2  &  5/2+ - 3/2-       &   119.44 & 120.5 & 0.17 &   $<$72.3\\

[\ion{N}{ii}] & $^3$P$_2$-$^3$P$_1$            &   121.90 & 188.20 &  0.12    &  45.7\\
CO     &         21-20     &   124.19 & 1276.10 & 0.17     &  357.1\\
p-H$_2$O & 4$_{04}$-3$_{13}$ & 125.35 & 319.5 & 0.12 & 42.8 \\
CO     &         20-19     &   130.37 & 1160.24 & 0.16     &  319.0\\
CO     &         19-18     &   137.20 & 1049.89 & 0.18    &  338.7\\
p-H$_2$O  &  3$_{13}$-2$_{02}$&  138.53 & 204.7 & 0.17    &  103.0\\
CO     &         18-17     &   144.78 & 945.01 & 0.15    &  350.4\\

[\ion{O}{i}] & $^3$P$_0$-$^3$P$_2$   &   145.53 & 326.58 & 0.16     &   146.7\\
CO     &         17-16     &   153.27 & 845.63 & 0.15    &  386.0\\
CO     &         16-15     &   162.81 & 751.76 & 0.18    &  466.5\\
OH 1/2-1/2  &   3/2+ - 1/2- &   163.12 & 270.2 & 0.24     &  91.0\\
OH 1/2-1/2  &  3/2+ - 1/2-  &   163.40 & 269.8 & 0.19     &   90.0\\
CO       &        15-14    &   173.63 & 663.37 & 0.16    &  413.0\\
o-H$_2$O    &3$_{03}$-2$_{12}$&   174.63 & 196.8 & 0.19     &  208.0\\
o-H$_2$O    &2$_{12}$-1$_{01}$&   179.53 & 114.4 & 0.11     &   31.0\\
o-H$_2$O    &2$_{21}$-2$_{12}$ & 180.49 & 194.1 & 0.12 & 34.9 \\
CO       &       14-13     &   186.00 & 580.51 &  0.20    &  437.2\\   \hline
\end{tabular}\\ 
\end{center}
\label{tab:lines}
$^a$ Laboratory wavelength. 
$^b$ Flux summed over all pixels in 10$^{-18}$ W m$^{-2}$. Typical line flux uncertainties are dominated by calibration uncertainties ($\sim$ 30-50$\%$) and fringing (0.1 Jy at the longest wavelengths to 0.6 Jy at the shortest). \\
$^1$ Lines blended with above line. 
$^2$ Line detected only in central spaxel and should not be compared to the fluxes of the summed spaxels.
$^3$ Doublet unresolved.
\end{table}
}

\def\placeTableCont{
\begin{table}[!th]
\caption{Best-fit from Robitaille models}
\begin{center}
\begin{tabular}{l l | l l }
\hline \hline
\multicolumn{4}{c}{DK Cha characteristics} \\ \hline
Envelope Mass & 0.03 M$_\odot$ & A$_V^1$ & 7.9 \\
Disk Mass & 0.03 M$_\odot$ & Inclination & $<18^{\circ}$\\
Luminosity & 29.4 L$_\odot$ & Disk Inner Radius & 0.3 AU\\ \hline

\end{tabular}\\
\end{center}
$^1$ Total A$_{\rm{v}}$ of which A$_{\rm{v}}$=5.2 mag. is attributed to the envelope at an inclination of 18$^\odot$ and 2.7 mag. to more extended cloud.
\label{tab:cont}
\end{table}
}


\abstract 
{}
{DK Cha is an intermediate-mass star in transition from an embedded
configuration to a star plus disk stage.
We aim to study the
composition and energetics of the circumstellar material during this
pivotal stage.
}
{Using the Range Scan mode of PACS  on the Herschel Space Observatory, 
we obtained a spectrum of DK Cha from 55 to 210 $\mu$m as part of the DIGIT Key Program.
}
{Almost 50 molecular and atomic lines were detected, many more than the 7
lines detected in ISO-LWS. Nearly the entire ladder of CO from $J$=14-13 to
38-37 ($E_u/k = 4080$ K), water from levels as excited as $J_{K_{-1}K_{+1}} =
7_{07}$ ($E_u/k = 843$ K), and OH lines up to $E_u/k = 290$ K were detected.
}
{The
continuum emission in our PACS SED scan matches the flux expected from
a model consisting of a star, a surrounding disk of 0.03 M$_\odot$, and an
envelope of a similar mass, supporting the suggestion that the object is
emerging from its main accretion stage. Molecular, atomic, and ionic
emission lines in the far-infrared reveal the outflow's
influence on the envelope.
The inferred hot gas can be photon-heated, but some 
emission could be due to C-shocks in the walls of the outflow cavity. 
}


\maketitle

\section{Introduction}

One of the
 pivotal points in the formation of a star occurs when the
 star and disk system emerges from the collapsing envelope, signaling
 the end of the main accretion of stellar mass.
The transition from Stage I protostar, embedded in a substantial envelope, to
a Stage II configuration, with only a star and disk\footnote{See, e.g., \citet[][]{Robitaille07} for a definition of the Stages}
is poorly understood.

IRAS 12496-7650 is the brightest infrared (IR) and submillimeter
(submm) source in the Chamaeleon II star-forming region, at a distance of
$178 \pm 18$ pc \citep{Whittet97}.
Follow-up studies to the discovery of IRAS 12496-7650 by
\citet{Hughes89,Hughes91}, along with subsequent observations
\citep[e.g.,][]{Whittet97,Young05a,Porras07,Spezzi08},
identified the variable Herbig Ae star DK Cha as the 
optical counterpart.
The central star mass is estimated at
2-3 M$_\odot$ \citep{Spezzi08}, with a
bolometric luminosity between 24 \citep{Spezzi08}
and 35 L$_\odot$ \citep{vanKempen09c}. 
The star is surrounded by a molecular outflow and protostellar envelope 
\citep{vanKempen06,vanKempen09c}.
The spectral energy distribution (SED) is roughly flat in wavelength
from near-infrared to mid-infrared \citep{Hughes89}. 
Using the $T_{\rm{bol}}$ definition and including longer wavelength data,
\citep[e.g.][]{Henning93,vanKempen09c}, DK Cha has $T_{\rm{bol}} = 580$ K,
making it a Class I source, but near the Class II boundary at 650K.
Due to an approximately face-on viewing angle down the outflow cone
\citep{vanKempen09c}; one can observe the protostellar disk directly,
unobstructed by the surrounding envelope. This viewing angle 
also allows the optical emission to escape.
The observational SED Classes are often assumed to correspond to the 
physical Stages,
but this is by no means obvious when aspherical models are examined
\citep{Robitaille07,Crapsi08,Dunham10} or tracers of dense gas are considered
\citep{vanKempen09}. The same physical object can have different SED 
classifications depending on viewing angle.
Thus, it is interesting to study an object that is near
an SED classification boundary and that also
appears to be transitional in physical stage.

Strong far-IR continuum and line emission have been detected
\citep{Giannini99,vanKempen06}.
As discussed by \citet{vanDishoeck04,Nisini05}, and
\citet{vanKempen10H} for the case of the similar source HH46,
far-IR atomic, ionic, and molecular lines can constrain the properties
of the disk, envelope, and outflow.
CO emission lines between $J$=14--13 and $J$=19--18, as well as [\ion{O}{i}]
and [\ion{C}{ii}], were observed from DK Cha with ISO-LWS, but other possible molecular
species, such as OH and H$_2$O, were undetected
\citep{Lorenzetti99,Giannini99}.

\placeTablePACS

The large beam of ISO-LWS (90$''$) and its low spectral
resolving power ($\lambda/\Delta\lambda = 200$)
were insufficient to determine the
origin (e.g., envelope versus outflow) of the CO emission. The preferred
picture was of a small, quiescent hot core of $\sim$300 AU around the
protostar \citep{Giannini99}, but recent observations of CO 4--3
and 7--6 indicate that this warm CO emission
either originates outside the inner few thousand AU or
cannot be heated simply by radiation from the central star
(passive heating) \citep{vanKempen06};
the far-IR and submm lines have different energies and
critical densities, so they may originate in different regions.

\placeFigureSED

The Key Programme, Dust, Ice, and Gas in Time (DIGIT) aims to study the
evolution of these three constituents through the star formation process
by observing objects in different stages. This paper, along with
a companion paper on HD100546, a Stage II object \citep{Sturm10H},
present the first results from the DIGIT Science Demonstration observations.

\section{Observations}

The launch of the Herschel Space Observatory \citep{Pilbratt10}, with the Photodetector Array
Camera and Spectrometer (PACS) instrument \citep{Poglitsch10}, enables far-IR spectra
with high spectral resolving power and good spatial ($\sim$9$''$) resolution.
The range scan/SED mode provides a full spectrum (55-210 $\mu$m). 

DK~Cha was observed for a total of $\sim$4 hours on 10 December 2009
by the PACS spectrometer on the Herschel Space Observatory
in pointed range-scan spectroscopy mode
(obsid 1342188039 and 1342188040).  An image slicer rearranges the
light into a $5\times5$ array, with each spatial pixel (spaxel) covering
$9\farcs4\times9\farcs4$.  A grating disperses the light onto two
16 $\times$ 25 pixel Ge:Ga detectors. Observations were
taken using nodding, in which two off-positions separated by $\pm 6$ arcmin in
elevation were taken as $`$blank$'$ sky, from which the signal is subtracted.
Consequently, spatially diffuse lines (e.g., [\ion{C}{ii}]) are 
subtracted out.

The effective spectral resolving power was about 1000+/-500, lower than was expected. This is due to the pointing error, which is perpendicular to the slit, as well as due to fringing issues on the continuum.  The point-spread
function of PACS increases towards longer wavelengths and is larger
than a single spaxel, but is corrected in post-processing using the PACS
spectrum of the pointlike HD100546 \citep{Sturm10H} to calculate the extraction
efficiencies.

\placeFigureLines

The data were reduced using the Herschel Interactive Processing Environment
(HIPE v. 2.4). The observation was slightly mispointed ($\sim$5-6$''$) 
making it hard to assess whether emission is extended and causing 
wavelength shifts in some spectral ranges. Because of the
mispointing, uncertainty in the Spectral Response Function, light
leakage problems, and uncertainties in the wavelength dependence of the PSF,
the spectrum was trimmed shortward of 56 $\mu$m, longward of 190 $\mu$m,
and from 95-101 $\mu$m. Two spectra were used to identify lines.
One is from the central spaxel alone. For a second spectrum, used for
most lines, fluxes from all spatial pixels were summed to correct 
for the PSF and mispointing.  We will focus on gas lines

\section{Results}

\subsection{Continuum}

The total Spectral Energy Distribution (SED) in Fig. 1 includes photometry
from the literature
\citep[Nefs et al., in prep]{Alcala08, Spezzi08, Evans09}, combined with
Spitzer-IRS spectroscopy and the PACS results.
The photometric points were fitted by the SED fitter routine 
from \citet{Robitaille07}, with
results for parameters in Table \ref{tab:cont}. Due to the extinction and variability at short wavelengths, only points longwards of 3 $\mu$m were taken as input for the fit.
The best-fit model is overplotted and split into the contributions from the
star, disk, and envelope.  
The short-wavelength IRS continuum is dominated by the flux from the disk, the
longer wavelength IRS continuum covers the transition from disk to envelope,
and the PACS continuum arises  in the envelope.
The model underestimates the observations in the near-IR and mid-IR,
indicating that the disk emission is underestimated in the model.


The best-fit models of the SED grid indeed confirm the assumption of a
nearly face-on configuration $<$18$^{\circ}$, with relatively low and equal
envelope and disk masses of 0.03 M$_\odot$, indicative of a source in transition
from Stage I to Stage II \citep{Robitaille07}.
The PACS continuum was not used to constrain the fit;
the consistency with the predicted model (differences
of $<$10$\%$) is thus encouraging, especially considering
the calibration uncertainty of 30-50$\%$.

\placeTableCont

\subsection{Line Inventory}

Table \ref{tab:lines} lists all gas lines that have been identified.
Fig. 2 shows the 
spectrum after subtraction of a polynomial fit to the continuum.
We used a combination of the ISO line
list\footnote{http://www.mpe-garching.mpg.de/iso/linelists/Molecular.html}
and the list published in \citet{Lerate06} for the Orion bar.
More lines may be detectable in this spectrum after
PSF deconvolution. Some lines in Table \ref{tab:lines} are only
detected in the central spaxel. Due to the fringing in other spaxels at short wavelengths and the PSF problems discussed above, these do
not show up in the summed result of Fig. 2.
Using spectra both from the summation of all 25 spaxels and
from only the central spaxel, a large range of CO, H$_2$O
and OH lines were detected, ranging in excitation energies from a 
few hundred kelvin to over 3000 K. In addition, [\ion{O}{i}] and [\ion{N}{ii}]
were identified. Emission from [\ion{C}{ii}] at 157.7 $\mu$m is  not 
seen. That line was detected by \citet{Lorenzetti99}, but was believed to be extended emission, associated with the galactic ISM. This component is
 subtracted out in our observations due to the nodding, 
which was not used in ISO-LWS.

\placeFigureCO

\subsection{Gas analysis}

Far-infrared CO lines were previously detected by \citet{Giannini99},
but only up to CO $J$=19-18. Assuming the lines to be optically thin, those could be
fitted with temperatures ranging from $T=200$ K  to $T=750$ K 
and densities from $n=4 \times 10^6$ cm$^{-3}$ 
to $n=5 \times 10^5$ cm$^{-3}$. The high T, low n model, which
produces the most emission in the high-$J$ lines, is 
overplotted in Fig. 3.
As can be seen, this model severely underproduces the CO lines from higher
transitions seen by Herschel. 
The other proposed models produce even less high-$J$ emission.

The CO level populations up to $J=31$ (Fig. 3), can be fitted with two
distinct rotational temperatures (\trot),
with the break at about $E_{\rm{up}}=1500$ K.
Most of the mid-J CO lines can be fitted
by $\trot \sim 380$ K, while higher excitation lines are best fitted with
$\trot \sim 1365$ K. 
In HH46 \citep{vanKempen10H}, 
UV photon heating \citep{Spaans95} is dominant for lines with energies 
below 1700 K, while a non-dissociative C-shock dominates the emission 
from higher excitation lines. A similar model is plausible for DK Cha,
and the issue will be explored
in a forthcoming paper (Green et al. in prep).

Water was detected in a wide variety of transitions,
up to 7$_{07}$-6$_{16}$ at 71.93 $\mu$m with ($E/k = 843$ K).
Water was not detected in ISO-LWS \citep{Giannini99} because the lines are
5 times weaker than the CO lines.
The discovery of the 7$_{07}$-6$_{16}$ line indicates high temperature,
as confirmed by the high-$J$ CO lines.  
Given the possible identification of shocked CO and the geometry of DK
Cha, it is likely that the water emission arises at least in part from 
either a non-dissociative C-shock or UV-heating of the quiescent 
material along the outflow cavity walls.

In addition, at least 9 lines of OH are detected.
Since oxygen is rapidly transformed into H$_2$O at temperatures
higher than 100 K  and frozen out below that, either a very strong UV field
or a dissociative shock is needed.  OH is a major coolant for dissociative
shocks \citep{Neufeld89}. This dissociative shock is accompanied
by a C-shock, which may explain the CO and H$_2$O analysis.
Models from Wampfler et
al. (in prep) also suggest that C-shock densities are often too low to both
account for these high excitation lines and at the same time not overproduce
the 119 and 163 $\mu$m OH lines. 

HD 100546 (Sturm et al. 2010), a Herbig Be star without an envelope,
was chosen for early observations to be a source analogous to DK Cha, 
but further evolved. 
A comparison is useful because similar species are seen in its spectrum. 
Emission of H$_2$O and OH  from the upper layers of a flared, but quiescent
disk, similar to that around HD 100546 \citep{Sturm10H}, 
could account from some of the lines from DK Cha. However, accounting
for the difference in distance, the CO lines from DK Cha are more
than 10 times stronger, and higher excitation lines from H$_2$O are
seen toward DK Cha. 
A possible explanation for the increased emission of DK Cha is 
an accretion shock onto the disk, where high densities and
temperatures are present. Another possibility is a shock in the
walls of the outflow. This will also be further examined 
in Green et al. in prep.

\section{Conclusions}

Continuum models that fit the data are consistent with the interpretation of 
DK Cha
as transitional between a Stage I and Stage II object, with nearly equal
masses in envelope and disk and a nearly face-on geometry.
The full PACS SED mode observations reveal an extremely rich spectrum of
over 50 ionic, atomic, and molecular lines.  With further improvements to the
spectral response function, we will be able to 
identify additional gas, ice and dust features in the spectrum.
The full range scan/SED mode can probe many aspects of a protostellar
object, from the ambient envelope to gas excited by ultraviolet radiation
and by shocks.

\begin{acknowledgements}

Support for this work, part of the Herschel Open Time Key Project Program,
was provided by NASA through an award issued by the Jet Propulsion Laboratory,
California Institute of Technology.
The authors would like to thank Bart VandenBussche and Alessandra Contursi for help with the
data reduction, and David Ardila for his help in scheduling the observations
prior to the ESAC SDP workshop.

\end{acknowledgements}
\bibliographystyle{./aa}
\bibliography{./biblio}
\end{document}